\documentclass[pre,aps,amsmath,twocolumn,showpacs,superscriptaddress,
longbibliography]{revtex4-1}

\usepackage{soul} % para tachar palabras: \textst{normal}

\usepackage{amssymb}
\usepackage{graphicx}
\usepackage[pdftex,bookmarks,colorlinks,breaklinks]{hyperref}
\hypersetup{linkcolor=red,citecolor=blue,filecolor=dullmagenta,urlcolor=blue}
\usepackage{amsmath}
\usepackage{mathrsfs}
\usepackage{amsfonts}

\begin{document}

\title{Universal properties of Wigner delay times and resonance widths of tight-binding random graphs}

\author{K.~B. Hidalgo-Castro}
%\email{blass.khic@gmail.com}
\affiliation{Instituto de F\'isica, Benem\'erita Universidad Aut\'onoma de Puebla, Puebla 72570, Mexico}

\author{L.~A. Razo-L\'opez}
%\email{luis-alberto.razo-lopez@univ-cotedazur.fr}
\affiliation{Universit\'e C\^ote d'Azur, CNRS, Institut de Physique de Nice (INPHYNI), Nice, France}
\affiliation{Institut Langevin, ESPCI Paris, Universit\'e PSL, CNRS, 75005 Paris, France}

\author{A.~M. Mart\'inez-Arg\"uello}
\email{blitzkriegheinkel@gmail.com}
\affiliation{Instituto de F\'isica, Benem\'erita Universidad Aut\'onoma de Puebla, Puebla 72570, Mexico}

\author{J.~A. M\'endez-Berm\'udez}
%\email{jmendezb@ifuap.buap.mx}
\affiliation{Instituto de F\'isica, Benem\'erita Universidad Aut\'onoma de Puebla, Puebla 72570, Mexico}
\affiliation{Escuela de F\'isica, Facultad de Ciencias, Universidad Nacional Aut\'onoma de Honduras, Honduras}

%====================================================================++

\begin{abstract}

The delay experienced by a probe due to interactions with a scattering media is highly related to the internal dynamics inside that media. This property is well captured by the Wigner delay time and the resonance widths. By the use of  the equivalence between the adjacency matrix of a random graph and the tight-binding Hamiltonian of the corresponding electronic media, the scattering matrix approach to electronic transport is used to compute Wigner delay times and resonance widths of Erd\"os-R\'enyi graphs and random geometric graphs, including bipartite random geometric graphs. In particular, the situation when a single-channel lead attached to the graphs is considered. Our results show a smooth crossover towards universality as the graphs become complete. We also introduce a parameter $\xi$, depending on the graph average degree $\langle k \rangle$ and graph size $N$, that scales the distributions of both Wigner delay times and resonance widths; highlighting the universal character of both distributions. Specifically, $\xi = \langle k \rangle N^{-\alpha}$ where $\alpha$ is graph-model dependent.

\end{abstract}


\maketitle

%====================================================================++

\section{Introduction}

Random graphs have attracted a lot of attention in the last decades as they have proved to be valuable models for describing complex systems with applications found in a broad variety of areas of research~\cite{Strogatz2001,Dorogovtsev2001,Albert2002,Newman2003,Boccalettia2006,DurretBook,LatoraBook,EstradaBook,PenroseBook,Erdos1959,Cimini2019,Ozkanlar2014,Dorogovtsev2008,Estrada2012,Cimini2019,Barabasi1999}. Up to date, many aspects that characterize the behavior of random graphs have been addressed. Those include connectivity and structural properties~\cite{Strogatz2001,Dorogovtsev2001,Albert2002,Newman2003,Boccalettia2006,DurretBook,LatoraBook,EstradaBook,PenroseBook}, percolation behavior~\cite{Callaway2000}, and spectral and eigenfuction statistical properties~\cite{Goh2001,Rodgers2005,Georgeot2010,Jalan2011,Farkas2002,Dorogovtsev2003}. Based on an electrical analogy, when friction is present in the graphs, further insight into the behavior of random graphs has also been gained by studying the transport properties such as the electrical conductance established between a source node and a sink node assigned to positive and zero potentials respectively~\cite{Lopez2005,Xue2010}. Those properties strongly depend on the graph degree distribution.

Another possibility to characterize the behavior of random graphs has been opened due to the equivalence established between the adjacency matrix and the tight-binding Hamiltonian matrix describing respectively a random graph and an electronic random media, where both matrices are usually represented by sparse random matrices~\cite{Martinez2013,AMMA2025,Aguilar2020}. Consequently, this equivalence has also motivated the use of random-matrix theory (RMT) methods and techniques to study spectral and scattering and transport properties of random graphs which in turn provide further information about the structure of the graphs~\citep{Martinez2013,AMMA2025,Aguilar2020,Sade2005,Bandyopadhyay2007,Jalan2007,Jalan2010}. In the latter, the opening of the systems is usually performed by attaching the graphs to the outside by means of perfectly conducting leads, so the fundamental quantity characterizing the scattering processes within the random media is the scattering matrix; or the $S$-matrix~\cite{Beenakker1997}. It is in this context, for instance, that an isolated-to-metallic crossover as well as universal behavior, well described by RMT, have been observed in the transport properties of Erd\"os-R\'enyi random graphs~\cite{Martinez2013} and random geometric graphs~\cite{AMMA2025}.

Following the aforementioned equivalence, it is the purpose of the present work to deepen in the understanding of scattering processes in random graphs by studying a transport property of special interest, namely, the delay experienced by an electronic wave due to interactions with a scattering media~\cite{Wigner1955,Smith1960}. That delay is highly related to the internal dynamics inside the media and it is well captured by the so-called Wigner delay time and the resonance widths which represent, respectively, the typical time an electronic wave remains in the scattering media and the lifetime an electronic wave remains in the corresponding resonant state due to the opening of the system. Furthermore, although these quantities have been of great interest in the realm of complex scattering~\cite{Fyodorov1997a,Fyodorov1997b,AMMA2017,Fyodorov2012,Novaes2022,Chen2021a,Chen2021b} (see also the review on time delays and resonance widths in Ref.~\cite{Fyodorov2011}) they also hold significant importance from an experimental viewpoint as they offer alternative means to characterize systems using scattering measurements, thereby circumventing the need for direct access to wave functions~\cite{Kottos2002,JAMB2005,AMMA2023}.

The graph models to be considered below correspond to Erd\"os-R\'enyi random graphs and random geometric graphs, including bipartite random geometric graphs. Opposite to Erd\"os-R\'enyi graphs, in which nodes are not embedded in a given space, in random geometric graphs nodes are distributed in a geometric space (Euclidian in most cases). Also, random geometric graphs have been introduced as models more suited to describe those real world situations in which a given system is constrained by geometrical spatial bounds such as urban planing~\cite{Zhong2014}, species-habitat networks, 5G wireless networks~\cite{Stegehuis2022}, and many others; see e.g.~\cite{Barthelemy2011} and references therein. Moreover, since a microwave realization of a 2D tight-binding scattering setup is available, our results for random geometric graphs could be straightforwardly verified experimentally in artificial photonic systems~\cite{Kuhl2010,Bittner2010,Bellec2013,Barkhofen2013,Aubry2020}.

The paper is organized as follows. In the next section, the graph models as well as the scattering setup (when the graphs support one-open channel) are introduced. The perfect coupling regime between the graphs and the lead, the scattering phases, the Wigner delay time, and the resonance widths are defined in Sect.~\ref{sec:Properties}. The numerical results are presented in Sect.~\ref{sec:Results}. Finally, the conclusions are given in Sect.~\ref{sec:Conclusions}.

%====================================================================++

\section{Graph models and scattering setup}
\label{sec:Models}

In this work, the focus is paid on the properties of scattering phases, Wigner delay times, and resonance widths of two graph models, namely, Erd\"os-R\'enyi graphs (ERGs) and Random Geometric Graphs (RGGs), including a variation of RGGs: Bipartite Random Geometric Graphs (BRGGs). On the one hand, ERGs are composed of $N$ independent vertices, or nodes, connected to each other with probability $p$, where $p\in [0, 1]$. That is, when $p = 0$ the ERGs consist of $N$ isolated nodes, while when $p = 1$, the graphs are fully connected. On the other hand, RGGs consist of $N$ vertices uniformly and independently distributed in the unit square. Two vertices are connected by an edge if their Euclidean distance is less than or equal to the connection radius $r\in [0, \sqrt{2}]$~\cite{PenroseBook,Dall2002}. Then, RGGs depend on the parameter pair $(N,r)$. Meanwhile, BRGGs are composed by $N$ vertices grouped into two disjoint sets, namely, set A with $s$ vertices and set B with $N-s$ vertices, $s\in [1, N-1]$, where there are no adjacent vertices within the same set. The vertices belonging to both sets are uniformly and independently distributed in the unit square, and two vertices are connected by an edge if their Euclidean distance is less or equal than the connection radius $r\in [0, \sqrt{2}]$~\cite{Stegehuis2022}. That is, BRGGs depend on the parameters $(N,s,r)$. Notice that, given BRGGs of total size $N$, the parameter $s$ defines the sizes of both sets A and B; hereafter, those graphs will be referred to as BRGGs($s$).

The randomly weighted versions of ERGs, RGGs, and BRGGs can be described by the following sparse tight-binding Hamiltonian
\begin{equation}
H = \sum_{n = 1}^{N} h_{nn} | n \rangle \langle n | + \sum_{n, m} h_{nm} ( | n \rangle \langle m | + | m \rangle \langle n | ) ,
\label{eq:H}
\end{equation}
where $N$ is the number of vertices in the graphs, $h_{nn}$ are on-site potentials, and $h_{nm}$ are hopping integrals between sites $n$ and $m$. In Eq.~(\ref{eq:H}), the second summation runs over all pairs of connected nodes. Furthermore, in Eq.~(\ref{eq:H}) the weights $h_{nm}$ are chosen to be statistically independent random variables drawn from a normal distribution with zero mean $\langle h_{nn} \rangle = 0$ and variance $\langle |h_{nm}|^{2} \rangle = (1 + \delta_{nm})/2$. In addition, each graph is assumed to be undirected i.e., $h_{nm} = h_{mn}$ such that $H=H^{T}$, with $T$ the transpose operation. From the physical point of view, those weights, $h_{nn}$ and $h_{nm}$, may represent respectively random self-loops and random strength interactions between vertices $n$ and $m$.

% FIGURE 1
%
\begin{figure}
\centering
\includegraphics[width=1.0\columnwidth]{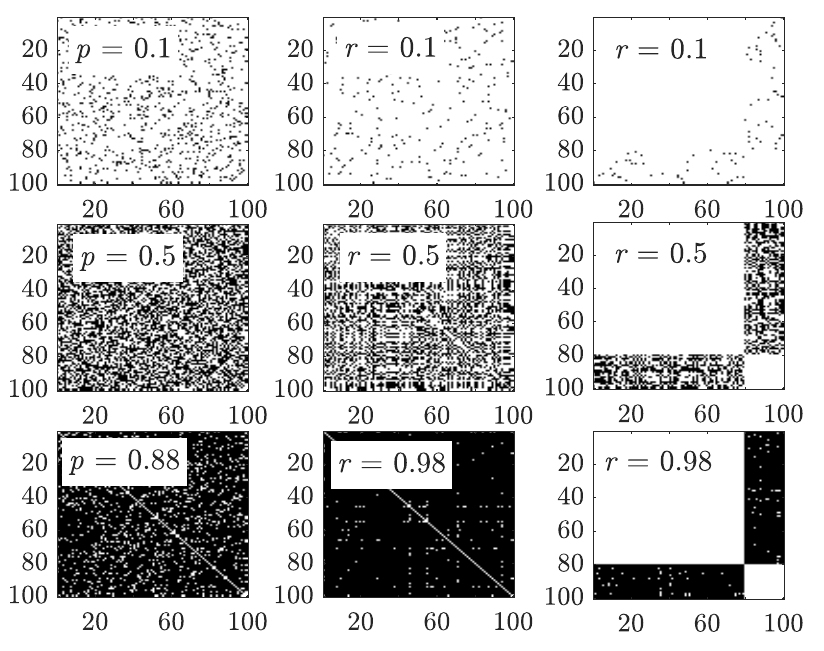}
\caption{Structure of Hamiltonian~(\ref{eq:H}) for ERGs (first column), RGGs (second column), and BRGGs (third column). Three values of connection probability $p$ and connection radius $r$ are considered. For the BRGGs case, $s = 4N/5$. }
\label{fig:HRGGs}
\end{figure}

As an example, in Fig.~\ref{fig:HRGGs} the structure of Hamiltonian~(\ref{eq:H}) is depicted for ERGs, RGGs, and BRGGs, in first, second, and third column, respectively. Different values of connection probability $p$ and connection radius $r$ are considered, as shown in the insets. The block structure of the Hamiltonian of BRGGs with $s=4N/5$ is clearly different to those of ERGGs and RGGs.

In the one open channel situation, the graphs are open to the outside by attaching to them a single-channel lead described by the following 1D tight-binding Hamiltonian
\begin{equation}
H_{\mathrm{lead}} = \sum_{n=1}^{-\infty} (| n \rangle \langle n + 1 | + | n + 1 \rangle \langle n |) .
\end{equation}
In this case, due to the flux conservation condition, the scattering matrix, $S(E)$, reduces to a phase which can be written as~\cite{Beenakker1997,Verbaarschot1985}
\begin{equation}
S(E) = 
\mathrm{e}^{\mathrm{i} \phi(E)} =
1 - 2\mathrm{i} \sin(k) W^{T} \frac{1}{E - \mathcal{H}_{\mathrm{eff}} } W ,
\label{eq:S}
\end{equation}
where $k = \arccos(E/2)$ is the wave vector supported in the lead, $E$ is the energy, and $\mathcal{H}_\mathrm{eff}$ is the non-Hermitian effective Hamiltonian given by
\begin{equation}
H_\mathrm{eff} = H - \mathrm{e}^{\mathrm{i}k} W W^{T}
\label{eq:Heff}
\end{equation}
with $H$ the $N\times N$ tight-binding Hamiltonian matrix of Eq.~(\ref{eq:H}) that describes the isolated graphs with $N$ resonant states. In Eq.~(\ref{eq:S}), $W$ is an $N\times 1$ energy independent matrix that couples the $N$ resonant states to the single propagating mode in the lead. The elements of $W$ are equal to zero or $\varepsilon$, where $\varepsilon$ is the coupling strength between the graphs and the lead. Furthermore, assuming that the wave vector $k$ undergoes slight changes near the energy band center, we set $E = 0$ and neglect the energy dependence of $H_{\mathrm{eff}}$ and of $S$.

Since for ERGs and RGGs all the vertices are statistically equivalent, the single channel lead is attached to any vertex chosen at random. For BRGGs, however, as mentioned above, the vertices belong to two different disjoint sets, set A with $s$ vertices and set B with $N - s$ vertices, that is, vertices are not equivalent and thus there are several possibilities to attach the single channel lead. If $s = N/2$, for instance, both sets have the same number of vertices and the set where the lead is attached is not relevant. In previous studies~\cite{AMMA2025}, it has been shown that several scattering and transport properties of weighted BRGGs($s$) with $s = N/2$ and $N/5$ show good agreement with RMT predictions, then in this work we will consider the case $s = 4N/5$ with the single channel lead attached to the larger set; a setup where the RMT limit is not attained~\cite{AMMA2025}.

%====================================================================++

\section{scattering measures}
\label{sec:Properties}

In this section, the scattering measures used to analyze the properties of ERGs, RGGs, and BRGGs are presented. These correspond to scattering phases, Wigner delay times, and resonance widths when the graphs support one open channel. The analysis is performed in the so-called perfect coupling regime between the graphs and the lead. Other transport properties when random graphs support more than one open channel were analyzed, for instance, in Refs.~\cite{Martinez2013,AMMA2025}.

%====================================================================++

\subsection{Scattering phases and perfect coupling regime}

Within the framework of RMT and in the one-channel situation, that is, when a single-channel lead is attached to the system~(\ref{eq:H}), the scattering matrix reduces to a phase, $S(E) = \mathrm{e}^{\mathrm{i} \phi(E)}$ [see Eq.~(\ref{eq:S})], whose distribution is given by~\cite{Ossipov2005}
\begin{equation}
P(\phi) = \frac{1}{2\pi} \frac{1}{\gamma + \sqrt{\gamma^{2} - 1} \cos \phi} ,
\end{equation}
in the diffusive (metallic-like) regime, where $\gamma = (1 + |\langle S \rangle|^{2})/(1 - |\langle S \rangle|^{2})$ and $\langle S \rangle$ is the average over an ensemble of $S$ matrices. The averaged scattering matrix, $\langle S \rangle$, can be interpreted as the fraction of the incident waves which comes out promptly from the scattering region. Then, $\langle S \rangle$ provides a measure of the coupling strength between the system and external region~\cite{Feshbach1973,Brouwer1997,Fyodorov1997a}. In the perfect coupling regime $\langle S \rangle$ vanishes and the phase is uniformly distributed over the unit circle. Here, the analysis presented in next sections is performed in the perfect coupling regime between the graphs and the lead, that is, in the regime when $\langle S \rangle \approx 0$. This corresponds to the situation where none of the waves are reflected back promptly from the scattering region.

%====================================================================++

\subsection{Wigner delay time}
\label{sub:WignerDelayTime}

The delay experienced by a probe due to its interactions with a scattering region is well characterized by the so-called Wigner delay time $\tau$~\cite{Wigner1955,Smith1960}. It is defined in terms of the scattering matrix $S$ and its derivative with respect to the energy $E$. In the one channel case, near the center of the spectrum ($E =0 $), the Wigner delay time can be written as~\cite{Kottos2002,Steinbach2000}
\begin{equation}
\tau(E=0) = \left. \frac{\mathrm{d} \phi(E)}{\mathrm{d} E} \right|_{E=0} = -2\,\mathrm{Im}\, \mathrm{Tr}(E - H_{\mathrm{eff}})^{-1} |_{E=0}.
\end{equation}
Furthermore, the poles of the scattering matrix show up as resonances which correspond to the complex eigenvalues $\mathcal{E}_{n} = E_{n} - \mathrm{i}\Gamma/2$ of the effective non-Hermitian Hamiltonian $H_{\mathrm{eff}}$ [see Eq.~(\ref{eq:Heff})], where $E_{n}$ and $\Gamma_{n}$ are the position and width of the $n$th resonance, respectively. In addition, the lifetime of the $n$th resonance is related to the resonance width as $\tau_{n} = 1/\Gamma_{n}$.

In the one-channel situation and under the perfect coupling regime between the open channel and the scattering region, the Wigner delay time equals the partial delay time. A relation obtained between the negative moments of the delay time, scaled by $\Delta/2\pi$ with $\Delta$ the mean level spacing of the closed system, $\widetilde{\tau} = \frac{\Delta}{2\pi} \tau$, and that of the eigenfunction intensities also of the closed system, $\tilde{y} = V | \psi_{n}(\vec{r})|^{2}$ with $V$ the volume of the sample, leads to a functional relation among the distributions of both quantities~\cite{Ossipov2005}. Since in the RMT limit the distribution of $\tilde{y}$ is known, and given by the so-called $\chi^{2}$-distribution; for time-reversal symmetric systems the above mentioned relation is explicitly given by~\cite{Ossipov2005}
\begin{equation}
w(\widetilde{\tau}) = \frac{1}{\widetilde{\tau}^{3}} p\Big(\frac{1}{\widetilde{\tau}} \Big),
\label{eq:RelationB1}
\end{equation} 
where
\begin{equation}
p(\tilde{y}) = \frac{1}{\sqrt{2\pi}} \tilde{y}^{-1/2} \mathrm{exp}\Big(-\frac{\tilde{y}}{2}\Big)
\label{eq:pB1}
\end{equation}
is known as the Porter-Thomas distribution. Hence, the Wigner delay time distribution becomes
\begin{equation}
w(\widetilde{\tau}) = \frac{1}{\sqrt{2\pi}} \widetilde{\tau}^{-5/2} \mathrm{exp}\Big(-\frac{1}{2\widetilde{\tau}}\Big) ,
\label{eq:wB1}
\end{equation}
from which the average Wigner delay time is $\langle \widetilde{\tau} \rangle = \int_{0}^{\infty} \widetilde{\tau}\, w(\widetilde{\tau}) \mathrm{d} \widetilde{\tau} = 1$, i.e., $\langle \tau \rangle = \frac{2\pi}{\Delta}$. Now, since the mean level spacing $\Delta$ can be expressed in terms of the mean eigenvalue density $\rho(E)$ as $\Delta = \frac{1}{N \rho(E)}$, then~\cite{Fyodorov1997a}
\begin{equation}
\frac{\langle \tau \rangle}{N} = 2\pi\, \rho(E) ,
\label{eq:TauScaling}
\end{equation}
which is independent of the system size $N$. That is, for complete graphs, it is expected that $\frac{\langle \tau \rangle}{N}$ reflects the scaling dependence of $\rho(E)$ of the corresponding closed graphs, as will be shown below.

Furthermore, it is often convenient to analyze the distribution of the Wigner delay time~(\ref{eq:wB1}) in the transformed variables $x = \ln\, \tau^{-1}/\langle \tau^{-1} \rangle$ and $x = \ln\, \tau/\tau^{\mathrm{typ}}$. Those distributions can be obtained from distribution~(\ref{eq:wB1}) which take the forms
\begin{equation}
P\big(\ln\, \tau^{-1}/\langle \tau^{-1} \rangle\big) = \frac{1}{\sqrt{2\pi}}\, k_{1}^{-\frac{3}{2}}\, \mathrm{exp}\Big(\frac{3k_{1}x - \mathrm{e}^{x}}{2k_{1}}\Big)
\label{eq:DistInvTauAvg}
\end{equation}
and
\begin{equation}
P\big(\ln\, \tau/\tau^{\mathrm{typ}} \big) = \frac{1}{\sqrt{2\pi}}\, k_{2}^{-\frac{3}{2}}\, \mathrm{exp}\Big(-\frac{3k_{2}x + \mathrm{e}^{-x}}{2k_{2}}\Big),
\label{eq:DistTauTyp}
\end{equation}
where we have used the usual transformation $P(x) \mathrm{d} x = w(\widetilde{\tau}) \mathrm{d} \widetilde{\tau}$, and $k_{1} = \frac{\Delta}{2\pi}\, \frac{1}{\langle \tau^{-1} \rangle}$ and $k_{2} = \frac{\Delta}{2\pi}\, \tau^{\mathrm{typ}}$, respectively, and $\Delta$ as before is the mean level spacing of the closed counterpart.

%====================================================================++

\subsection{Resonance widths}
\label{sub:ResonanceWidths}

Besides the Wigner delay time, another quantity highly related to the dynamics of a probe within a scattering region is the well-known resonance width $\Gamma$, whose inverse represent the probe lifetime in a resonant state escaping into the open channel. For systems with complex dynamics, in the regime of RMT applicability, the distribution of resonance widths in the regime of weak coupling to the continuum is known~\cite{Fyodorov2012}. In addition, for time-reversal symmetric systems, in the regime of non-isolated resonances and in the multiple coupled channels case, the distribution of resonance widths have also been derived~\cite{Sommers1999,Fyodorov2022}. In the one channel situation and under the perfect coupling regime, it is obtained through~\cite{Fyodorov2022}
\begin{equation}
p(\tilde{z}) = \frac{1}{4 \sqrt{2}} \mathrm{e}^{-\tilde{z}} \mathbb{L}\, \phi(\tilde{z}) ,
\label{eq:pztilde}
\end{equation}
where $\tilde{z} = \frac{\pi}{\Delta} z$ and $\mathbb{L}$ is the operator
\begin{equation}
\mathbb{L} = 2 \sinh(2 \tilde{z}) - \Big[ \cosh(2 \tilde{z}) - \frac{\sinh(2 \tilde{z})}{2 \tilde{z}} \Big] \Big( \frac{3}{\tilde{z}} + 2 \frac{\mathrm{d}}{\mathrm{d} \tilde{z}} \Big)
\end{equation}
acting on the function
\begin{equation}
\phi(\tilde{z}) = \int_{1}^{\infty} \frac{a - 1}{\sqrt{a + 1}} \mathrm{e}^{-\tilde{z} a} \mathrm{d} a .
\end{equation}
Therefore, the scaled resonance width distribution~(\ref{eq:pztilde}) is explicitly given by
\begin{eqnarray}
p(\tilde{z}) & = & \frac{\mathrm{e}^{4\tilde{z}} - 4 \tilde{z} - 1}{8\, \tilde{z}^{2} \mathrm{e}^{4 \tilde{z}} } + \nonumber \\
& & \frac{\sqrt{2\pi}\, \mathrm{erfc}(\sqrt{2 \tilde{z}})\, \big[ \mathrm{e}^{4\tilde{z}} (4 \tilde{z} - 3) + 8 \tilde{z} (2 \tilde{z} + 1) + 3 \big]}{32\, \tilde{z}^{\frac{5}{2}} \mathrm{e}^{2 \tilde{z}} }, \nonumber \\
\label{eq:DistFy}
\end{eqnarray}
where $\mathrm{erfc}(z)$ is the complementary error function and in the changed variable $x = \ln \Gamma/\Gamma^{\mathrm{typ}}$, $p(\tilde{z})$ becomes
\begin{equation}
P(\ln \Gamma/\Gamma^{\mathrm{typ}}) = k_{3}\, \mathrm{e}^{x}\, p(k_{3}\, \mathrm{e}^{x}),
\label{eq:DistGGtyp}
\end{equation}
where $k_{3} = \frac{\pi}{\Delta} \Gamma^{\mathrm{typ}}$ and $p(z)$ is given by distribution~(\ref{eq:DistFy}).

Since spectral and transport statistical properties of complete time-reversal symmetric tight-binding random ERGs~\cite{Mirlin1991,Fyodorov1991,Fyodorov1997c}, RGGs, and BRGGs~\cite{AMMA2025}, follow GOE RMT behavior, it is reasonable to expect that distributions~(\ref{eq:DistInvTauAvg}), (\ref{eq:DistTauTyp}), and (\ref{eq:DistGGtyp}) are to be retrieved for complete graphs. As it will be shown below, this is indeed the case.

For the graph models considered in this work, due to the random configuration of the vertices and also due to the random strength of interactions between vertices, both quantities, Wigner delay times and resonance widths, show sample-to-sample fluctuations. Therefore, in what follows, a statistical analysis over an ensemble of equivalent graphs shall be considered.

%====================================================================++

\section{Results}
\label{sec:Results}

For the statistical analysis of Wigner delay times and resonance widths $\Gamma$ shown below, the calculations are performed near the center of the spectrum, i.e., close to $E \approx 0$ and under the perfect coupling regime between the graphs and the lead. In addition, for the statistics of $\Gamma$, we use an eigenvalue window of about one-tenth of the total spectra around $E \approx 0$. The graph sizes considered correspond to $N = 100$, 200, and 400, with random realizations, or ensemble sizes, such that the statistics is fixed to $10^{6}$ data. Also, the results displayed do not contain error bars since the statistics is performed with a large amount of data such that the error bars are smaller than the symbol size.

Prior to moving forward to the statistical analysis of Wigner delay times and of resonance widths, the perfect coupling condition between the graphs and the lead is first established. This condition is attained at the value of the coupling strength $\varepsilon = \varepsilon_{0}$ such that $\langle S(\varepsilon_{0}) \rangle \approx 0$. In this scenario, none of the waves are reflected back before entering the scattering region and, as a consequence, the internal structure of the scattering region can be probed.

In previous studies, a detailed numerical analysis of $\langle S \rangle$ as a function of $\varepsilon$ and the average degree $\langle k \rangle$ shows that the coupling strength given by~\cite{Martinez2013,AMMA2025}
\begin{equation}
\varepsilon_{0} \approx \left\{
\begin{array}{rl}
0.462\, \langle k \rangle^{0.29} + 0.374 & \text{for } \mathrm{ERGs}, \\
0.462\, \langle k \rangle^{0.29} + 0.374 & \text{for } \mathrm{RGGs}, \\
0.760\, \langle k \rangle^{0.02} + 0.060 & \text{for } \mathrm{BRGGs}(4N/5),
\end{array} \right.
\label{eq:BGepsilon0}
\end{equation}
sets the perfect coupling regime for ERGs, RGGs, and BRGGs, respectively, as a function of $\langle k \rangle$. That relation is independent of the graph sizes and of the number of attached leads, and it will be used in all our calculations to set the perfect coupling condition. Moreover, a scaling analysis has been shown that the parameter $\xi \propto \langle k \rangle N^{-\alpha}$ scales well the spectral and structural properties of directed RGGs~\cite{Peralta2023}. This has been done by noting that a localized-to-delocalized transition in the spectral properties displays a shift as a function of $\langle k \rangle$ for different graph sizes $N$. This shift is characterized by the values of $\langle k \rangle$, $\langle k \rangle^{*}$, at which the averaged spectral properties reach half the transition. A plot of $\langle k \rangle^{*}$ as a function of $N$ shows a power-law behavior which suggests the proposed scaling parameter $\xi \propto \langle k \rangle N^{-\alpha}$~\cite{Peralta2023}. Equivalently, a localized-to-delocalized transition has also observed in the behavior of the scattering matrix elements of ERGs and RGGs with a shift in the curves for different graph sizes $N$~\cite{AMMA2025}, characterized by the same parameter $\langle k \rangle^{*}$ that follows a power-law behavior as function of $N$. Thus, here we also consider the parameter
\begin{equation}
\xi = \langle k \rangle N^{-\alpha}
\label{eq:xi}
\end{equation}
to analyze Wigner delay times and resonance widths. In fact, as it will be shown below, (\ref{eq:xi}) also scales the properties of Wigner delay times and resonance widths, highlighting their universal character. For tight-binding ERGs, RGGs, and BRGGs($4N/5$), the exponents $\alpha = 0.075 \pm 0.0029, 0.26 \pm 0.0185$, and $0.3429 \pm 0.0371$, respectively, were obtained~\cite{AMMA2025}. For BRGGs($s$), other values of $\alpha$ can be obtained depending on the sizes of sets A and B. 

% FIGURE 2
%
\begin{figure}
\centering
\includegraphics[width=0.95\columnwidth]{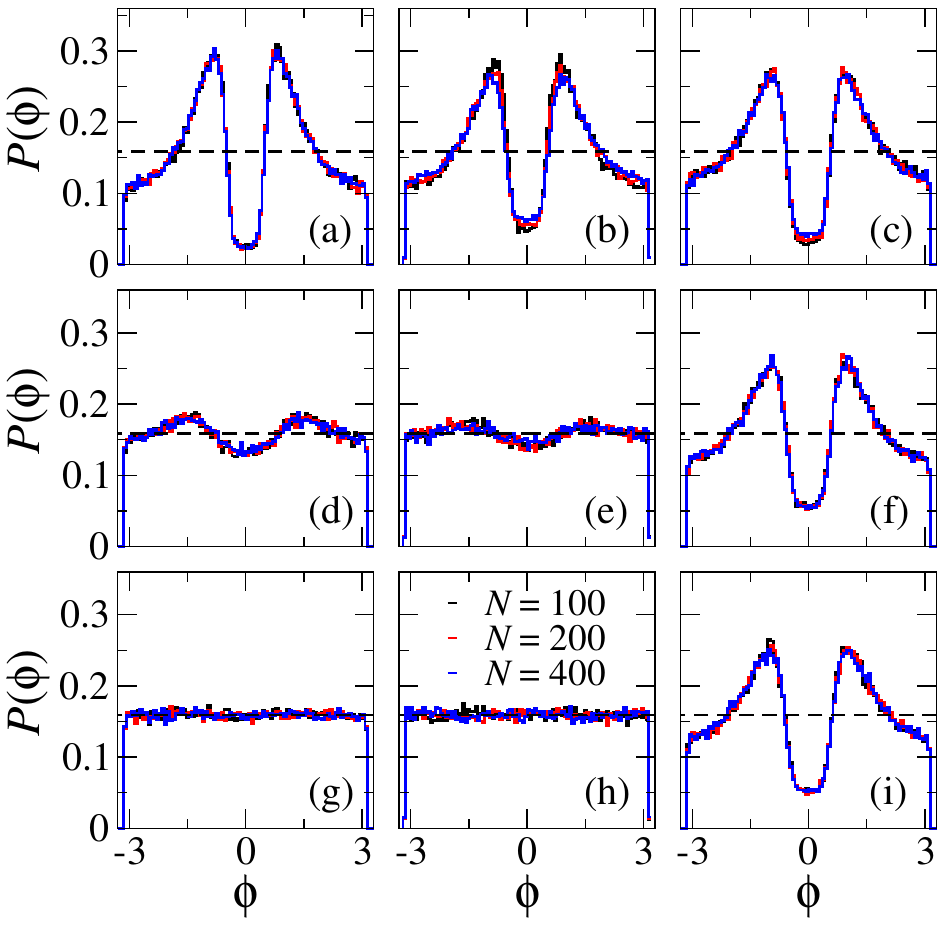}
\caption{Phase distribution in the perfect coupling regime for tight-binding ERGs (first column), RGGs (second column), and BRGGs($4N/5$) (third column). The graph sizes correspond to $N = 100$, 200, and 400, in black, red, and blue histograms. The scaling parameter is fixed to $\xi = 0.2\, (0.2, 0.2), 5.0\, (4.0, 1.8)$, and 70.08 (29.87, 6.59) for ERGs (RGGs, BRGGs) shown in upper, middle, and lower panels. These values lead to nearly isolated, half connected, and mostly connected graphs, respectively. Horizontal black dashed lines correspond to the uniform distribution $P(\phi) = 1/2\pi$.}
\label{fig:DistPhase}
\end{figure}

In the metallic-like or diffusive regime and under the perfect coupling condition, $\langle S \rangle \approx 0$, the phase of the $S$-matrix is uniformly distributed around $1/2\pi$, see Eq.~(\ref{fig:DistPhase}). Therefore, a crossover towards uniformity is expected in the phase distribution of random graphs as a function of the average degree $\langle k \rangle$ or the scaling parameter $\xi$. This behavior is observed in Fig.~\ref{fig:DistPhase} where the phase distribution for tight-binding ERGs, RGGs, and BRGGs, in left, middle, and right columns, respectively, is depicted considering three graph sizes, see panel (h). In each panel, the coupling strength $\varepsilon_{0}$, such that $\langle S(\varepsilon_{0}) \rangle \approx 0$, is set for graphs with a scaling parameter $\xi$ that leads to nearly isolated, half connected, and mostly connected graphs, in upper, middle, and lower panels, respectively. Those values correspond to $\xi = 0.2\, (0.2, 0.2), 5.0\, (4.0, 1.8)$, and 70.08 (29.87, 6.59) for tight-binding ERGs (RGGs, BRGGs).

As observed in Fig.~\ref{fig:DistPhase}, the phase distribution scales well with $\xi$ and the graphs size dependence drops off. That is, once $\xi$ is fixed, $P(\phi)$ is also fixed. Also, in the metallic-like regime, i.e., for mostly connected graphs, the phase is uniformly distributed around $1/2\pi$ for ERGs and for RGGs, see Figs.~\ref{fig:DistPhase} (g) and (h). For the case of BRRGs, the perfect coupling regime is not attained due to the lack of complexity in the graphs and the asymmetry of the scattering setup here considered.

% FIGURE 3
%
\begin{figure}
\centering
\includegraphics[width=1.0\columnwidth]{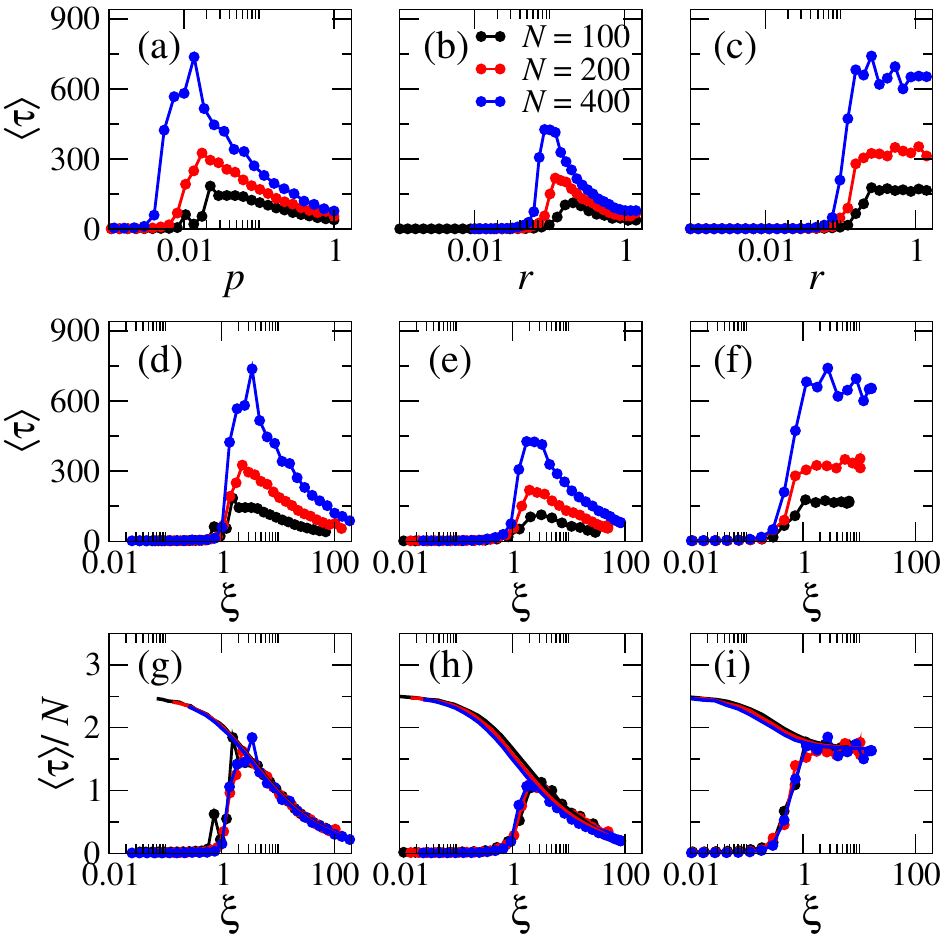}
\caption{Average Wigner delay time for tight-binding ERGs (first column), RGGs (second column), and BRGGs($4N/5$) (third column) as a function of the connection probability $p$ and the connection radius $r$ (upper panels), and as a function of scaling parameter $\xi$ (middle and lower panels). In panels (g-i), continuous lines are obtained from the closed graphs through the mean eigenvalue density $\rho$, see Eq.~(\ref{eq:TauScaling}). The graph sizes $N$ are indicated in panel (b). Each point is computed by averaging over $10^{6}$ random realizations.}
\label{fig:Tau}
\end{figure}

Once the perfect coupling condition is set, we now start our analysis by examining first the behavior of the Wigner delay time as a function of the parameters of the graph models under study. Figure~\ref{fig:Tau} shows the average Wigner delay time as a function of the connection probability $p$ and the connection radius $r$ in upper panels, and as a function of scaling parameter $\xi$ in middle and lower panels for ERGs, RGGs, and BRGGs, in first, second, and third column, respectively. The average is performed over $10^{6}$ realizations of random graphs. The graph sizes considered are indicated in panel (b). For nearly isolated graphs ($p\approx 0$ or $r\approx 0$), no time is spent by the electronic wave inside the media as shown in all panels of Fig.~\ref{fig:Tau}. As the graphs get more connected, the time spent inside the scattering media increases until it reaches a maximum for the connection probability $p \approx 0.01$, connection radius $r \approx 0.9$, and scaling parameter $\xi \approx 2.0$, respectively, as shown in each panel. Also, the larger the graph size $N$, the more time the wave spends inside the graphs, as observed in upper and middle panels. From their maximum time, the average Wigner delay time decreases as the graphs get mostly connected due to the appearance of multiple pathways. Furthermore, notice that even though the scaling parameter $\xi$ scales well several scattering and transport properties of tight-binding ERGs, RGGs, and BRGGs~\cite{AMMA2025}, this parameter does not scale well the average Wigner delay time, as shown in middle and lower panels of Fig.~\ref{fig:Tau}. In order to find a scaling of $\langle \tau \rangle$, the average Wigner delay time is normalized by the graph size, as it is shown in the lower panels of Fig.~\ref{fig:Tau} for the ERGs, RGGs, and BRGGs, in first, second, and third column, respectively. As observed, for the three graph sizes considered, the curves of $\langle \tau \rangle/N$ vs.~$\xi$ show a tendency to fall on top of each other where however, slight deviations are noticeable. In order to see the scaling behavior of $\langle \tau \rangle/N$, in Fig.~\ref{fig:Tau} we also compare $\langle \tau \rangle/N$ with the mean eigenvalue density $\rho(E)$ in continuous lines, see equation~(\ref{eq:TauScaling}), obtained from the corresponding closed graph of panels (g-i). As expected, the scaling by $N$ is suggested by the relation between $\langle \tau \rangle/N$ and $\rho(E)$ of the closed counterpart.

% FIGURE 4
%
\begin{figure}
\centering
\includegraphics[width=0.99\columnwidth]{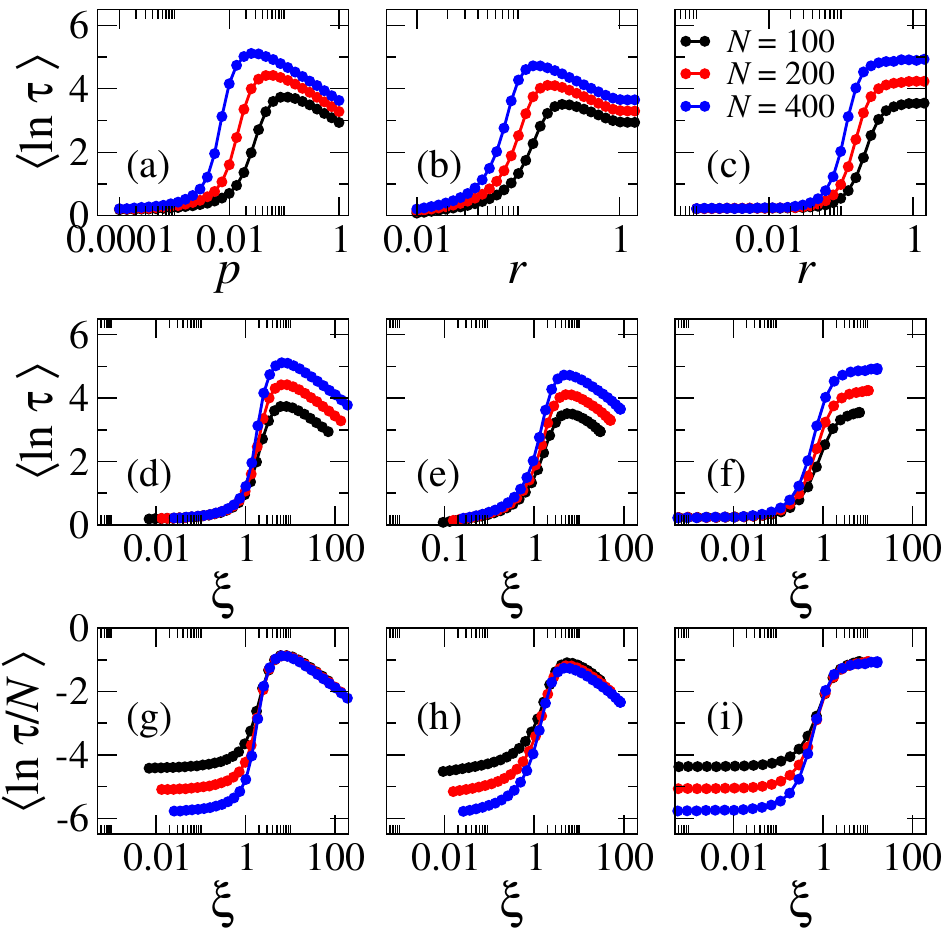}
\caption{Average of the logarithm of the Wigner delay time for tight-binding ERGs (first column), RGGs (second column), and BRGGs($4N/5$) (third column) as a function of connection probability $p$ and connection radius $r$ (upper panels), and as a function of scaling parameter $\xi$ (middle and lower panels). The results for three graphs sizes $N$ are reported: $N = 100$, 200, and 400, in black, red, and blue full circles, respectively. In every panel, each data value is computed by averaging over $10^{6}$ realizations of the corresponding random graph.}
\label{fig:LnTau}
\end{figure}

Also, further insight into the behavior of the Wigner delay time is provided by looking at its logarithm. For this purpose, the average of the logarithm of the Wigner delay time for ERGs, RGGs, and BRGGs, in first, second, and third column, respectively, as a function of $p$ and $r$ in upper panels, and as a function of scaling parameter $\xi$ in middle and lower panels, is reported in Fig.~\ref{fig:LnTau}. The graph sizes considered are indicated in the inset of panel (c). Since for very small values of $p$, $r$, or $\xi$ the graphs mostly consist of isolated nodes, then no diffusion occurs and no time is spent by the electronic wave inside the graphs. As the connectivity increases, the wave spends more time within the scattering region until a maximum time is reached for ERGs and RGGs. However, for BRGGs the time spent by the wave inside the media keeps increasing as a function of connection radius $r$ until its maximum is reached for complete bipartite graphs. As expected, the larger the graph size, the larger the time spent by the wave inside the media, as shown in panels (a-f). In order to scale this behavior, in middle and lower panels of Fig.~\ref{fig:LnTau}, $\langle \ln \tau \rangle$ and $\langle \ln (\tau/N) \rangle$ as a function of scaling parameter $\xi$, respectively, are depicted. As observed, $\xi$ scales quite well the behavior of $\langle \ln \tau \rangle$ and that of $\langle \ln (\tau/N) \rangle$ for values of $\xi < 1$ and $\xi > 3$, respectively, of ERGs and RGGs. For BRGGs, $\xi$ scales well the behavior of $\langle \ln (\tau/N) \rangle$ for values of $\xi > 1$. Even though, the parameter $\xi$ scales reasonably well the behavior of $\tau$, other relevant scattering and transport properties of tight-binding ERGs, RGGs, and BRGGs were shown to be properly scaled by $\xi$~\cite{AMMA2025}. 

% FIGURE 5
%
\begin{figure}
\centering
\includegraphics[width=0.99\columnwidth]{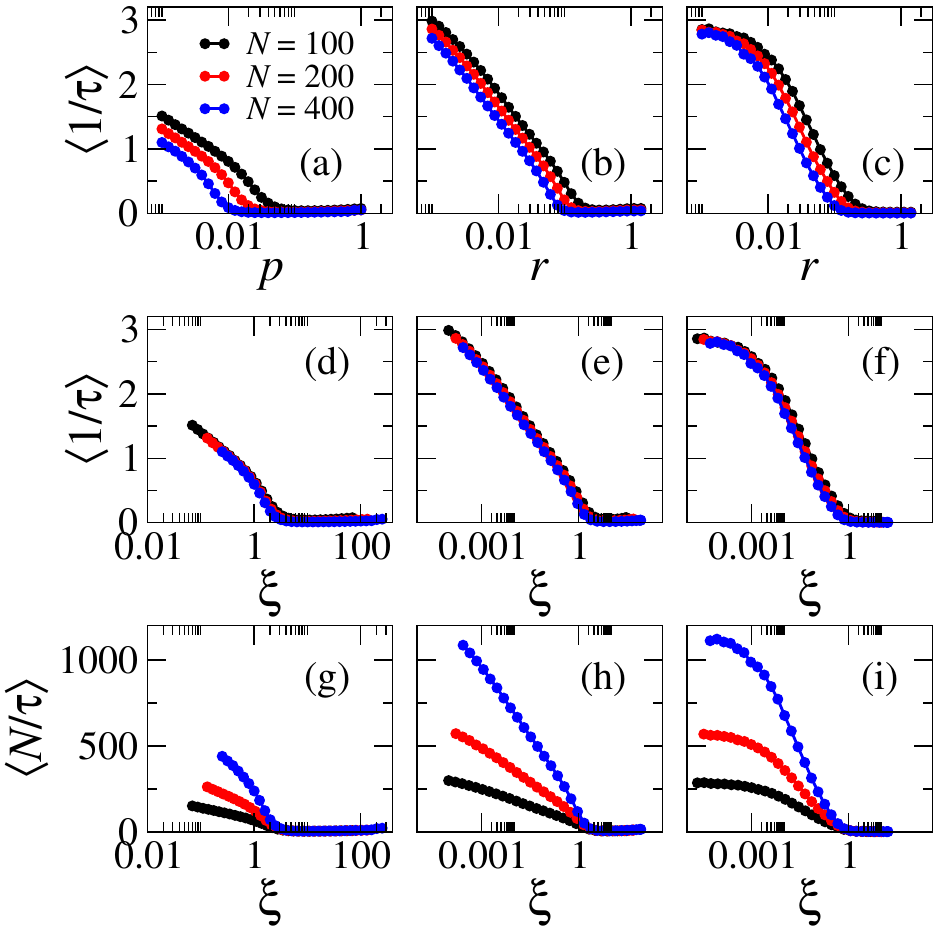}
\caption{Average of the inverse Wigner delay time for tight-binding ERGs (first column), RGGs (second column), and BRGGs($4N/5$) (third column) as a function of connection probability $p$ and connection radius $r$ (upper panels), and as a function of the scaling parameter $\xi$ (middle and lower panels). Three graphs sizes $N$ are considered: $N = 100$, 200, and 400. In all panels, each data value is computed by averaging over $10^{6}$ realizations of the corresponding random graph.}
\label{fig:InvTau}
\end{figure}

Moreover, as mentioned above, in the one channel situation both the partial delay times and the proper delay times (defined as the energy derivative of phase shifts and in terms of the scattering matrix and its derivative with respect to the energy, respectively) are identical to the Wigner delay time. Since the joint distribution of the reciprocals of the proper delay times is known and given by the Laguerre ensemble~\cite{Brouwer1997b}, then it is instructive to study the behavior of the inverse of Wigner delay time as well. Therefore, in Fig.~\ref{fig:InvTau} the average of the inverse Wigner delay time for ERGs, RGGs, and BRGGs, in first, second, and third column, respectively, as a function of $p$ and $r$ in upper panels, and as a function of $\xi$ in middle and lower panels, is reported. The graph sizes are indicated in panel (a). In every panel, each point is computed by averaging over $10^{6}$ realizations of the corresponding random graphs.
Upper panels of Fig.~\ref{fig:InvTau} show that as $p$ or $r$ increases, the average of the inverse Wigner delay time decreases, i.e., the time spent by the electronic wave inside the scattering media increases until it reaches its maximum (minimum of $\langle 1/\tau \rangle$). This indicates the localization of the electronic wave where conduction is suppressed. Also, the larger the graph size $N$, the faster the decay of the average inverse delay time. The graph size dependence is almost absent when plotting $\langle 1/\tau \rangle$ as a function of $\xi$ for ERGs, RGGs, and BRGGs for values of $\xi < 1$ with slight deviations noticeable for $\xi > 1$, see panels (d-f). Now, notice that a better scaling is obtained for values of $\xi > 1$ when plotting $\langle N/\tau \rangle$ vs $\xi$ for the three graph models under consideration, as shown in panels (g), (h), and (i) for ERGs, RGGs, and BRGGs, respectively. A similar panorama is also observed for different bipartitions of BRGGs($s$), for instance when $s = N/2$ and $N/5$ (not shown here).

%
% FIGURE 6
%
\begin{figure}
\centering
\includegraphics[width=0.99\columnwidth]{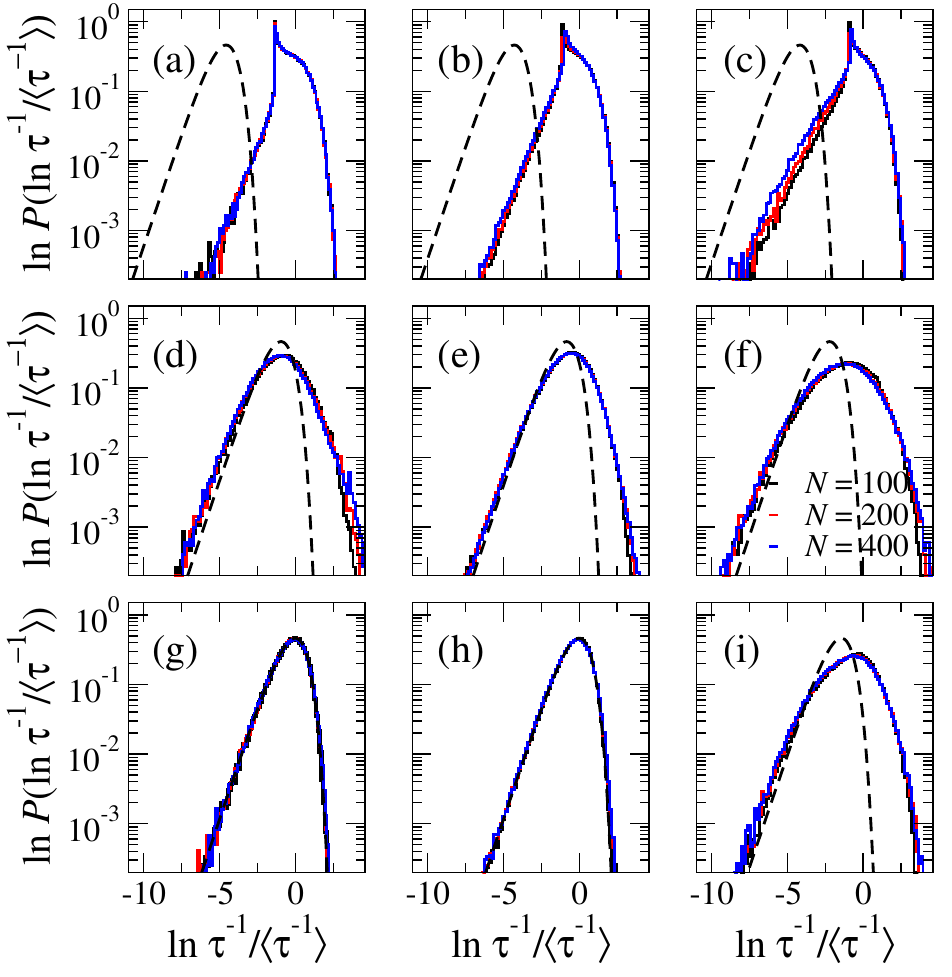}
\caption{Distribution of the logarithm of the inverse Wigner delay time for tight-binding ERGs (first column), RGGs (second column), and BRGGs($4N/5$) (third column). For the numerical results graphs sizes $N = 100$, 200, and 400, depicted in black, red, and blue, are considered. In upper, middle, and lower panels, the scaling parameter is fixed to $\xi = 0.2\, (0.2, 0.2), 5.0\, (4.0, 1.8)$, and 70.08 (29.87, 6.59) for ERGs (RGGs, BRGGs), respectively. These values of $\xi$ lead to graphs in the insulator, intermediate, and metallic regimes, respectively. Dashed lines correspond to distribution~(\ref{eq:DistInvTauAvg}) where the parameters of the graphs in each case are given in Table~\ref{tab:Table1}. As observed in panels (g) and (h), excellent agreement between numerical data and distribution~(\ref{eq:DistInvTauAvg}) is obtained.}
\label{fig:DistInvTau}
\end{figure}
%
%

%
%  TABLE 1
\begin{table*}

\vspace{-0.5cm}
{\caption{\label{tab:Table1} Values of the graph parameters to calculate $k_{1,2}$ used in distributions~(\ref{eq:DistInvTauAvg}) and (\ref{eq:DistTauTyp}). For all cases, the graph sizes considered are $N = 100$.}}
\vspace{-0.3cm}

\begin{center}
\begin{tabular}{cccccc} \hline \hline
\multicolumn{3}{c}{}    
                                            & \hspace{-0.2cm} ERGs & \hspace{0.2cm} RGGs & \hspace{0.4cm} BRGGs($4N/5$) \\ \hline
 
$\xi$                                       &  &  & \hspace{0.2cm} 0.2 \hspace{0.25cm} 5.0 \hspace{0.35cm} 70.08   
                                                  & \hspace{0.2cm} 0.2 \hspace{0.55cm} 4.0 \hspace{0.35cm} 29.87  
                                                  & \hspace{0.3cm} 0.2 \hspace{0.6cm} 1.8 \hspace{0.4cm} 6.59   \\
                                         
$\Delta \times 10^{-2}$                     &  &  & \hspace{-0.0cm} 2.6 \hspace{0.3cm} 4.8 \hspace{0.4cm} 15.9   
                                                  & \hspace{0.2cm} 2.8 \hspace{0.5cm} 6.1 \hspace{0.4cm} 15.9
                                                  & \hspace{0.3cm} 2.9 \hspace{0.65cm} 3.6 \hspace{0.5cm} 3.7   \\
                                                                     
$\langle \tau^{-1} \rangle \times 10^{-2}$  &  &  & \hspace{-0.2cm} 120.5 \hspace{0.15cm} 5.9 \hspace{0.45cm} 7.6  
                                                  & \hspace{0.1cm} 96.3 \hspace{0.35cm} 6.7 \hspace{0.5cm} 7.7
                                                  & \hspace{0.2cm} 89.1 \hspace{0.4cm} 15.6 \hspace{0.4cm} 7.5   \\
                                                                     
$\tau^{\mathrm{typ}}$                       &  &  & \hspace{-0.0cm} 1.38 \hspace{0.2cm} 39.54 \hspace{0.15cm} 18.95  
                                                  & \hspace{0.2cm} 1.61 \hspace{0.25cm} 31.02 \hspace{0.15cm} 18.96
                                                  & \hspace{0.4cm} 1.78 \hspace{0.3cm} 22.22 \hspace{0.1cm} 34.74    \\
                                   
$k_{1} \times 10^{-3}$                      &  &  & \hspace{0.1cm} 3.0 \hspace{0.2cm} 129.0 \hspace{0.15cm} 332.0  
                                                  & \hspace{0.2cm} 4.0 \hspace{0.4cm} 145.0 \hspace{0.2cm} 329.0
                                                  & \hspace{0.2cm} 5112.0 \hspace{0.2cm} 36.0 \hspace{0.2cm} 79.0    \\
                                   
$k_{2} \times 10^{-3}$                      &  &  & \hspace{0.1cm} 5.0 \hspace{0.2cm} 301.0 \hspace{0.1cm} 479.0  
                                                  & \hspace{0.2cm} 7.0 \hspace{0.4cm} 301.0 \hspace{0.2cm} 480.0
                                                  & \hspace{0.3cm} 8108.0 \hspace{0.1cm} 126.0 \hspace{0.1cm} 206.0    \\ \hline \hline
\end{tabular}

\end{center}

\end{table*}

In what follows, we turn our attention on the distribution of the inverse Wigner delay time and of the Wigner delay time normalized to its typical value. As it is reported below, both distributions show universal behavior for graphs in a wide range of transport regimes when scaled by the parameter $\xi$. Figure~\ref{fig:DistInvTau} shows the distribution of the logarithm of the inverse Wigner delay time for tight-binding ERGs (first column), RGGs (second column), and BRGGs($4N/5$) (third column), considering three graph sizes as indicated in panel (f). In upper, middle, and lower panels, the scaling parameter $\xi$ is fixed to values that lead to graphs into the insulator, in between the insulator-to-metallic, and in the metallic regimes, respectively. Those values correspond to $\xi = 0.2\, (0.2, 0.2), 5.0\, (4.0, 1.8)$, and 70.08 (29.87, 6.59) for tight-binding ERGs (RGGs, BRGGs), respectively. As observed, there exists a slight dependence on the graph size in the distribution of $\ln P(\ln \tau^{-1})$ for very low connected graphs, or for graphs in the insulator regime; particularly for BRGGs; see panel (c). This dependency is diminished as the graphs get more connected.

For comparison purposes, in Fig.~\ref{fig:DistInvTau} we have also included the corresponding theoretical distribution~(\ref{eq:DistInvTauAvg}) (black dashed lines) which depends on the parameters describing each graph for each value of $\xi$ through $k_{1} = \frac{\Delta}{2\pi} \frac{1}{\langle \tau^{-1}\rangle}$. Those values are given in Table~\ref{tab:Table1}. As observed, deviations between theory and numerics appear for graphs in the insulator and in between insulator-to-metallic regimes; see panels (a-f). As expected, for complete graphs an excellent agreement between the numerical data and distribution~(\ref{eq:DistInvTauAvg}) is obtained, see panels (g) and (h). For BRGGs($4N/5$), let us recall that the vertices belong to two different sets where adjacent vertices within the same set are not allowed. In this case, the single channel lead is attached to the larger set where the incoming waves are not able to explore homogeneously the full graph and, as a consequence, the metallic (RMT) regime is not reached and then deviations between numerics and distribution~(\ref{eq:DistInvTauAvg}) appear, see panel (i) of Fig.~\ref{fig:DistInvTau}. Other cases, for instance BRGGs($N/2$) or BRGGs($N/5$), with the single channel lead attached to the smaller set reaches the metallic regime where scattering and transport properties are described well by RMT measures~\cite{AMMA2025}.

Furthermore, let us note that after rescaling by the parameter $\xi$, the curves for $\ln P(\ln \tau^{-1})$ fall on top of each other not only for different graph sizes, but also for different connectivities $\xi$ and graph models under consideration, even though a slight dependence on graph size is observed in panel (c) of Fig.~\ref{fig:DistInvTau}. As connectivity increases, however, this size dependency diminishes. This collapse in the distribution of Wigner delay time is an indication of universal behavior shared by the three graph models analyzed. For the closed counterpart, this kind of universality has also been observed in the behavior of topological and spectral properties of RGGs~\cite{Aguilar2020}. Moreover, a crossover to RMT universal behavior is also observed where a perfect agreement with the Wigner delay time distribution~(\ref{eq:DistInvTauAvg}) is obtained. As observed, this trend to universal behavior is attained from half-connected to mostly connected graphs, as shown in middle and lower panels of Fig.~\ref{fig:DistInvTau}.

%
% FIGURE 7
%
\begin{figure}
\centering
\includegraphics[width=0.96\columnwidth]{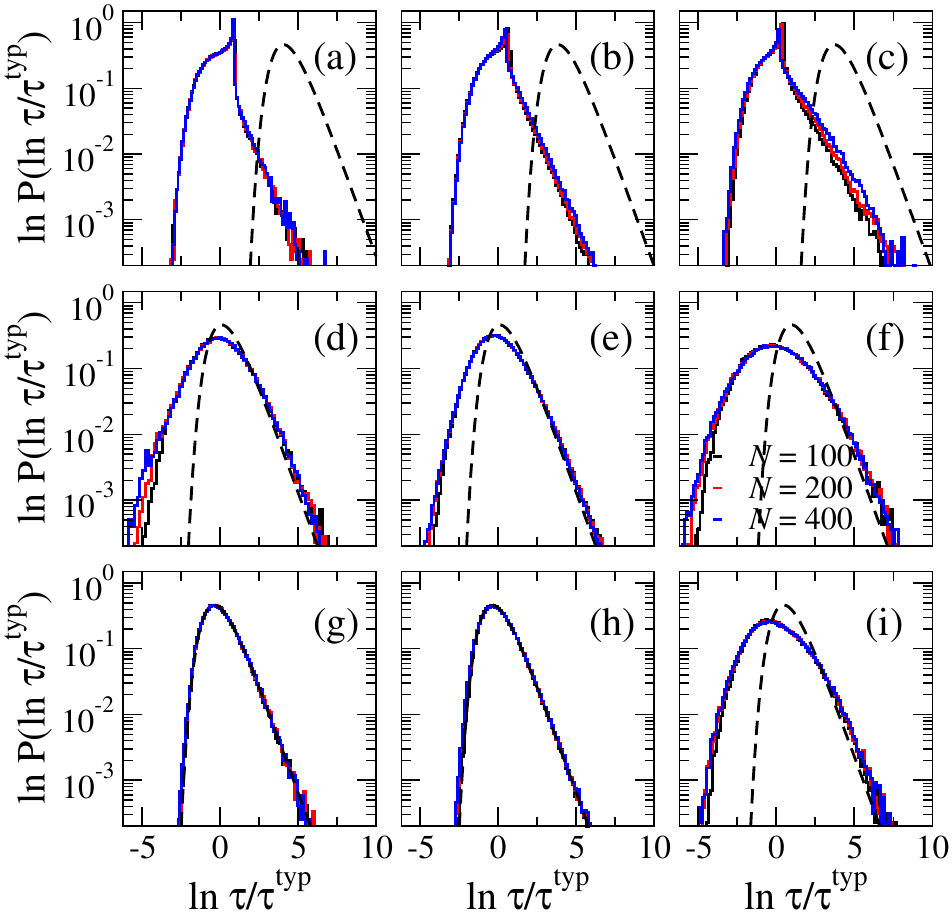}
\caption{Distribution of the logarithm of the Wigner delay time, normalized to its typical value, as a function of $\ln( \tau/\tau^{\mathrm{typ}})$ for tight-binding ERGs (first column), RGGs (second column), and BRGGs($4N/5$) (third column). The numerical results are obtained considering graphs sizes $N = 100$, 200, and 400, depicted in black, red, and blue histograms. In upper, middle, and lower panels the scaling parameter is fixed to $\xi = 0.2\, (0.2, 0.2), 5.0\, (4.0, 1.8)$, and 70.08 (29.87, 6.59) for tight-binding ERGs (RGGs, BRGGs), respectively. These values of $\xi$ lead to graphs in the insulator, intermediate, and metallic regimes, respectively. Dashed lines correspond to distribution~(\ref{eq:DistTauTyp}) where the parameters of the graphs in each case are given in Table~\ref{tab:Table1}. For complete graphs, excellent agreement between numerical data and distribution~(\ref{eq:DistTauTyp}) is obtained, see panels (g) and (h).}
\label{fig:DistTauTyp}
\end{figure}
%

% TABLE 2
%
\begin{table*}

\vspace{-0.5cm}
{\caption{\label{tab:Table2} Parameter values obtained to calculate $k_{3}$ used in distribution~(\ref{eq:DistGGtyp}). For all cases, the graph sizes considered are $N = 100$.}}
\vspace{-0.3cm}

\begin{center}
\begin{tabular}{cccccc} \hline \hline 
\multicolumn{3}{c}{}    
                                               & \hspace{-0.8cm} ERGs & \hspace{-0.3cm} RGGs & \hspace{0.3cm} BRGGs($4N/5$) \\ \hline

$\xi$                                    &  &  & \hspace{-0.35cm} 1.5 \hspace{0.8cm} 5.0 \hspace{0.9cm} 70.08 
                                               & \hspace{0.05cm} 1.5 \hspace{0.75cm} 4.0 \hspace{0.85cm} 29.87 
                                               & \hspace{0.25cm} 0.8 \hspace{0.8cm} 2.0 \hspace{0.65cm} 6.59   \\
                                       
$\Delta \times 10^{-2}$                  &  &  & \hspace{-0.45cm} 3.34 \hspace{0.6cm} 4.77 \hspace{0.9cm} 15.91  
                                               & \hspace{-0.05cm} 4.23 \hspace{0.6cm} 6.08 \hspace{0.8cm} 15.91
                                               & \hspace{0.15cm} 3.53 \hspace{0.6cm} 3.60 \hspace{0.6cm} 3.72   \\

$\Gamma^{\mathrm{typ}} \times 10^{-8}$   &  &   & \hspace{-0.2cm} 175.9 \hspace{0.2cm} 212800.0 \hspace{0.2cm} 1564000.0
                                                & \hspace{0.25cm} 9.24 \quad 241800.0 \hspace{0.2cm} 1558000.0
                                                & \hspace{0.2cm} 523.7 \quad 49470.0 \hspace{0.2cm} 97640.0   \\
                                   
$k_{3} \times 10^{-6}$                   &  &  & \hspace{-0.35cm} 165.4 \hspace{0.2cm} 140100.0 \hspace{0.3cm} 308800.0  
                                               & \hspace{0.1cm} 6.86 \quad 124900.0 \hspace{0.2cm} 307600.0
                                               & \hspace{0.15cm} 466.0 \quad 43100.0 \hspace{0.2cm} 82400.0   \\ \hline \hline

\end{tabular}
\end{center}

\end{table*}

%--------  TABLE

%
% FIGURE 8
%
\begin{figure}
\centering
\includegraphics[width=0.99\columnwidth]{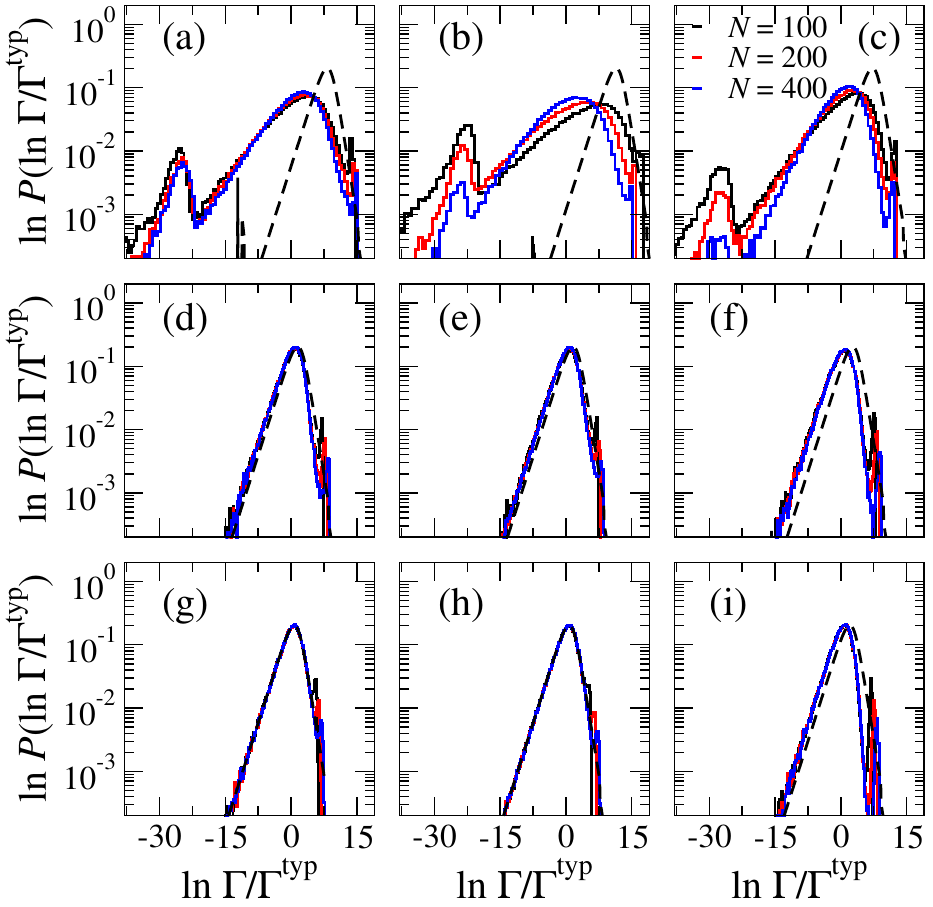}
\caption{$\ln P[\ln (\Gamma/\Gamma^{\mathrm{Typ}})]$ as a function of $\ln( \Gamma/\Gamma^{\mathrm{Typ}})$ for tight-binding ERGs (first column), RGGs (second column), and BRGGs($4N/5$) (third column). The numerical results correspond to graphs sizes $N = 100$, 200, and 400, depicted in black, red, and blue histograms. In upper, middle, and lower panels, the scaling parameter is fixed to $\xi = 1.5\, (1.5, 0.8), 5.0\, (4.0, 2.0)$, and 70.08 (29.87, 6.59) for ERGs (RGGs, BRGGs), respectively, which lead to graphs in the insulator, in between the insulator-to-metallic, and in the metallic regimes. Dashed lines correspond to distribution~(\ref{eq:DistGGtyp}) where the parameters of the graphs in each case are given in Table~\ref{tab:Table2}. A tendency to universal behavior as the graphs get more connected is observed.
}
\label{fig:DistGGtyp}
\end{figure}

In an equivalent way, the tendency to universal behavior can also be observed in the distribution of the logarithm of the Wigner delay time, normalized to its typical value $\tau^{\mathrm{typ}} \equiv \mathrm{exp} \langle \ln \tau \rangle$. This behavior is reported in Fig.~\ref{fig:DistTauTyp} which shows $\ln P(\ln \tau/\tau^{\mathrm{typ}})$ as a function of $\ln( \tau/\tau^{\mathrm{typ}})$ for the three graph models: ERGs (first column), RGGs (second column), and BRGGs (third column). Also, dashed lines correspond to distribution~(\ref{eq:DistTauTyp}) which depends on the graphs parameters $\Delta$ and $\tau^{\mathrm{typ}}$  through $k_{2} = \frac{\Delta}{2\pi} \tau^{\mathrm{typ}}$ for each value of $\xi$. Those parameters are given in Table~\ref{tab:Table1}. Again, three graph sizes are considered as indicated in panel (f). As in Fig.~\ref{fig:DistInvTau}, the scaling parameter is fixed to $\xi = 0.2\, (0.2, 0.2), 5.0\, (4.0, 1.8)$, and 70.08 (29.87, 6.59) for tight-binding ERGs (RGGs, BRGGs), respectively, which leads to graphs in the insulator (upper panels), in between the insulator-to-metallic (middle panels), and in the metallic regimes (lower panels). As observed, all the curves fall on top of each other highlighting the universal character of $\ln P(\ln \tau/\tau^{\mathrm{typ}})$ as well. As expected, RMT universal behavior is also reached for complete graphs as shown in Fig.~\ref{fig:DistTauTyp}, panels (g) and (h), where numerical data shows perfect agreement with distribution~(\ref{eq:DistTauTyp}).

Finally, a quantity highly related to the Wigner delay time is the resonance width whose reciprocal describes the lifetime of a probe in a resonant state escaping into the open channel. This quantity has attracted a lot of attention in the context of complex scattering, see for instance the review~\cite{Fyodorov2011}. Therefore, in Fig.~\ref{fig:DistGGtyp} we report the distribution of the logarithm of the resonance widths $\ln P[\ln (\Gamma/\Gamma^{\mathrm{typ}})]$, normalized to its typical value $\Gamma^{\mathrm{typ}} \equiv \mathrm{exp} \langle \ln \Gamma \rangle$. Right, middle, and left columns show the results considering tight-binding ERGs, RGGs, and BRGGs, respectively, with the graph sizes indicated in panel (c). In upper, middle, and lower panels, the scaling parameter is fixed to $\xi = 1.5\, (1.5, 0.8), 5.0\, (4.0, 2.0)$, and 70.08 (29.87, 6.59) for ERGs (RGGs, BRGGs), respectively. Opposite to the cases previously considered, for the analysis of resonance widths, in upper panels the parameter $\xi$ is fixed to $\xi = 1.5$ ($1.5, 0.8$) to set the graphs nearly in the insulator regime. Those values were chosen in order to avoid numerical errors. The values of $\xi$ considered in middle and lower panels set the graphs in between the insulator-to-metallic and in the metallic regimes, respectively. In Fig.~\ref{fig:DistGGtyp}, we also included in dashed lines the theoretical distribution~(\ref{eq:DistGGtyp}) valid in the metallic (RMT) regime. In order to obtain $k_{3} = \frac{\pi}{\Delta} \Gamma^{\mathrm{typ}}$, required in distribution~(\ref{eq:DistGGtyp}), for each graph model with fixed value of $\xi$, the mean level spacing $\Delta$ of the closed graph and $\Gamma^{\mathrm{typ}}$ were numerically computed. For each case considered in Fig.~\ref{fig:DistGGtyp}, the values are given in Table~\ref{tab:Table2}.

Now, even though some deviations are noticeable for low connected graphs (see upper panels in Fig.~\ref{fig:DistGGtyp}), a tendency to universal behavior of $\ln P[\ln (\Gamma/\Gamma^{\mathrm{typ}})]$ is observed. As in the distribution of Wigner delay times, Figs.~\ref{fig:DistInvTau} and \ref{fig:DistTauTyp}, also the curves for the distribution of resonance width, rescaled by $\xi$, fall on top of each other for different graph sizes, connectivity $\xi$, and the graph models under consideration. This also indicates the universal behavior of the width distribution shared by the three graph models analyzed here. However, this is not the case for low connected graphs, where a dependence on graph size is observed (see upper panels) even though this dependency diminishes as the graphs get more connected. Again, a crossover to RMT universal behavior is also observed for fully connected ERGs and RGGs models where a perfect agreement with the resonance width distribution~(\ref{eq:DistGGtyp}) is obtained.  As mentioned above, in BRGGs($4N/5$) in the scattering setup where the single channel lead is attached to the larger set, the incoming waves do not explore homogeneously the full graph, then BRGGs($4N/5$) does not reach the RMT limit and thus deviations betweem numerical data and RMT distribution~(\ref{eq:DistGGtyp}) appear. This is observed in Fig.~\ref{fig:DistGGtyp}, panel (i). The peaks appearing on the right side of every curve in each panels of Fig.~\ref{fig:DistGGtyp} correspond to finite size effects, as they diminish as the graph size increases.

%====================================================================++

\section{Conclusions}
\label{sec:Conclusions}

From the scattering matrix approach to electronic transport and from the equivalence established between the adjacency matrix and the tight-binding Hamiltonian matrix describing, respectively, a random graph and an electronic media, a numerical analysis of the Wigner delay times and the resonance widths of tight-binding Erd\"os-R\'enyi random graphs, random geometric graphs, and bipartite random geometric graphs, has been presented. Wigner delay times and resonance widths are related to the delay experienced by a probe due to interactions with a scattering media and provide important information about the internal dynamics inside that media.
Our results show that the distribution of both quantities, Wigner delay times and resonance widths, for random graphs with a single channel lead attached perfectly to them, are invariant under a scaling parameter $\xi$ which depends on the average degree and the graph size; see Eq.~(\ref{eq:xi}). That is, universal behavior in the distribution of Wigner delay times and of resonance widths is observed when scaled by $\xi$. In addition, a crossover to RMT universal behavior in the distribution of Wigner delay times and of resonance widths where perfect agreement between numerical data and analitycal distributions~(\ref{eq:DistInvTauAvg}), (\ref{eq:DistTauTyp}), and (\ref{eq:DistGGtyp}) for mostly connected graphs is also observed.

Even though the graph models considered in this work do not posses critical properties in the sense of an Anderson transition (i.e.~here we observe a smooth crossover from localization to diffusion by increasing $\xi$), it is worth mentioning that such crossover can be well captured by the Wigner delay time, a transport property which is very convenient from an experimental point of view since direct access to the eigenfunctions of the scattering media is not required.

Also our results may be experimentally verified in photonic systems in the microwave regime which emulate tight-binding models. Specifically, the 2D photonic arrays reported in Refs.~\cite{Kuhl2010,Bittner2010,Aubry2020} can be seen as a photonic tight-binding realization of random geometric graphs.

%====================================================================++

\section*{Acknowledgments}

L.A.R.-L. gratefully acknowledges the financial support from CONAHCyT (Mexico), through the Grant No.~775585, and from the French government, through the UCA \textsuperscript{JEDI} Investments in the Future project managed by the National Research Agency (ANR) with the reference number ANR-15-IDEX-0001. A.M.M.-A. acknowledges financial support from CONAHCyT under the program ``Estancias Posdoctorales por M\'exico 2022".
J.A.M.-B. thanks support from VIEP-BUAP (Grant No.~100405811-VIEP2025), Mexico.

%====================================================================++

%====================================================================++

%\bibliography{thispaper_py}

%====================================================================++

\end{document}